# Theoretical bounds for the exponent in the empirical power-law advance-time curve for surface flow

Behzad Ghanbarian[1*], Hamed Ebrahimian[2], Allen G. Hunt[3,4], and M. Th. van Genuchten[5,6]

[1] Department of Geology, Kansas State University, Manhattan KS, USA

[2] Department of Irrigation and Reclamation Eng., University College of Agriculture and Natural Resources, University of Tehran, Karaj Iran

[3] Department of Physics, Wright State University, Dayton OH, USA

[4] Department of Earth and Environmental Science, Wright State University, Dayton OH, USA

[5] Multiscale Porous Media Lab, Department of Earth Sciences, Utrecht University, Utrecht, The Netherlands

[6] Department of Nuclear Engineering, Federal University of Rio de Janeiro, Rio de Janeiro, Brazil

[*] Corresponding author's email address: ghanbarian@ksu.edu

**Abstract**

A fundamental and widely applied concept used to study surface flow processes is the advance-time curve characterized by an empirical power law with an exponent $r$ and a numerical prefactor $p$ (i.e., $x = pt^r$). In the literature, different values of $r$ have been reported for various situations and types of surface irrigation. Invoking concepts from percolation theory, we related the exponent $r$ to the backbone fractal dimension $D_b$, whose value depends on two factors: dimensionality of the system (e.g., two or three dimensions) and percolation class (e.g., random or invasion percolation with/without trapping). We showed that the theoretical bounds of $D_b$ are in well agreement with experimental ranges of $r$ reported in the literature for two furrow and border irrigation systems. We also used the value of $D_b$ from the optimal path class of percolation theory to estimate the advance-time curves of four furrows and seven irrigation cycles. Excellent agreement was obtained between the estimated and observed curves.

**1. Introduction**

Modeling water flow and solute transport in surface irrigation systems requires knowledge of how quickly water moves over the soil-air interface along a furrow or border. Analyses of this type lead to the advance time curve, which describes the balance among water that already exists in soil (referred to as the initial water content), water that infiltrates into the soil, and water that traverses on the surface. One of the most widely used forms in the literature for the advance time curve is the following empirical power law

$$x = pt^r \tag{1}$$

in which $t$ is the time needed for the wetting front to reach distance $x$, $p$ is a numerical prefactor and $r$ is an empirical exponent. Different values of $r$ have been reported experimentally. For example, Elliot and Walker (1982) estimated the parameters $p$ and $r$ by directly fitting Eq. (1) to two measured data points of the advance-time curve. Such a procedure, known as the two-point method, was first used for furrow irrigation and later applied to basin and border irrigation systems. Shepard et al. (1993) fixed $r$ at 0.5 and optimized $p$ by using typically the last measured point on the advance-time curve. Their method is known as the one-point method. Valiantzas et al. (2001) later presented another one-point method. Although they did not fix the value of $r$, their method requires a trial and error procedure to determine $r$ and $p$ from the last point of the advance-time curve.

In addition to one- and two-point methods, the values of $p$ and $r$ can be determined by directly fitting Eq. (1) to advance-time data if measurements are available. For instance, Serralheiro (1995) estimated $r$ for a Mediterranean soil (southern Portugal) using furrows of different slopes, and found $0.52 \leq r \leq 0.94$ (see their Table 1). In another study, Alvarez (2003) reported $0.58 \leq r \leq 0.72$ for different types of soils (such as Gleysols, Vertisols, Acrisols, and Ferrasols) for furrows of lengths between 240 and 380 m, with Manning's resistance coefficient $n$ ranging from 0.02 to 0.04. In Table 1, we summarize the range of $r$ value reported in different studies under various conditions. Experimental evidence indicates clearly that $r$ is a function of several factors, such as soil type, initial water content, soil surface slope, and type of surface irrigation. Thus, one should expect $r$ to vary from one soil and/or an irrigation cycle to another.

The main objective of this study was to invoke concepts from percolation theory to shed light on the value of $r$, and relate it to the backbone fractal dimension $D_b$ that characterizes the backbone structure for percolation. In what follows, we first introduce percolation theory and present its fundamental concepts. We then propose a power law from percolation theory and show the link between the backbone fractal dimension $D_b$ and the advance-time curve exponent $r$ in Eq. (1). We next compare our theoretical results with experiments for furrow and border irrigation systems.

**2. Percolation theory**

Percolation theory, introduced in its present form by Broadbent and Hammersley (1957), provides a promising theoretical framework to study interconnectivity and its effects on flow properties in heterogeneous systems and networks, such as soils. Broadbent and Hammersley (1957) studied plant disease spreading in an orchard whose trees were located on the intersections of a square lattice. As expected, when the distance between aligned trees increases, the probability of spreading a disease decreases. Eventually the distance between trees would reach a critical value above which the disease cannot spread over the orchard (Feder, 1988).

After the initial study by Broadbent and Hammersley (1957), various types of percolation theory have been used to model different phenomena. These include (1) random or ordinary (Stauffer, 1985), (2) invasion (Wilkinson and Willemsen, 1983), (3) directed (Obukhov, 1980), and (4) gradient (Rosso et al., 1986) approaches. In this study, we, however, focus on random and invasion percolation types.

Percolation theory exists in three main forms: site, bond, and continuum. For simplicity, we introduce here only the bond percolation theory by providing an example, and refer interested readers to Stauffer (1985), Feder (1988), Sahimi (1994; 2011) and Hunt et al. (2014b) for further details and other percolation types and forms.

Let us consider a simple regular square lattice as shown in Fig. 1a, in which only 10% of the bonds (the thick black segments) are occupied randomly and independently of the occupation status of their neighbors. The probability $p$ of a bond being occupied by a black segment is 0.1 in Fig. 1a. In this case, several small clusters consisting of one or two bonds exist. As the bond occupation probability $p$ (i.e., the fraction of bonds that are occupied) increases, more individual clusters become connected to each other and form a larger interconnected cluster (see Fig. 1b). The critical probability $p_c$ for bond percolation on the square network is 0.5 (Stauffer, 1985). In Fig. 1c, in which $p = 0.6$ (i.e., above the critical probability $p_c$), a large cluster connects the right side of the square network to its left side, and the top to the bottom. This means that percolation across the network has occurred.

A fundamental concept in percolation theory is the existence of a percolation threshold $p_c$, below which a system (or network) loses its connectivity (see Fig. 1a, b). For $p < p_c$, all clusters are finite in size, with the largest clusters having a typical size of the order of the finite correlation length ($\chi$), the mean distance between any two bonds on the same finite cluster, while no spanning cluster exists (Kirkpatrick, 1973; Ben-Avraham and Havlin, 2000). At $p_c$, an incipient infinite cluster occurs along with other finite clusters (Kirkpatrick, 1973; Ben-Avraham and Havlin, 2000). In this case, the mass of the spanning cluster increases with the size, $L$, of the lattice as a power-law $L^{D_f}$ in which $D_f$ is the mass fractal dimension of the fractal cluster whose universal value is 91/48 (or 1.896) and 2.53 in two and three dimensions, respectively (Feder, 1988). When $p > p_c$, the incipient infinite cluster still exists besides the finite clusters (Kirkpatrick, 1973).

At or below the percolation threshold, the typical size of the largest finite clusters is on the order of the finite correlation length. However, the infinite cluster above $p_c$ is different from the one at $p_c$ (Ben-Avraham and Havlin, 2000). When the system size $L$ is less than the correlation length (i.e., $L < \chi$), the system is a heterogeneous, statistically self-similar fractal (Gefen et al., 1983; Sahimi, 1993). For $L > \chi$ the system is macroscopically homogeneous, and the geometry is Euclidean (Feder, 1988; Sahimi, 1993).

Figure 2a shows the same square lattice as in Fig. 1c, with a bond occupation probability of 0.6. As can be seen, dead ends exist that do not contribute to flow. If such dead ends are removed from the cluster, what remains is called the backbone as shown in Fig. 2b. The backbone has a large number of loops, thus making it a multiply-connected object that contribute to transport.

Lee et al. (1999) studied traveling time for tracer particles and modeling flow driven by a pressure difference between two points separated by Euclidean distance $x$. They demonstrated that the most probable traveling time, $t$, of particles on a percolation cluster is proportional to $x^{D_b}$. Therefore,

$$x \propto t^{1/D_b} \tag{2}$$

where $D_b$ is the fractal dimension of the backbone. Equation (2) and its subsequent derivations have been evaluated successfully in their ability to describe the weathering rate of porous media (Hunt et al., 2014a; Hunt, 2015; Yu and Hunt, 2017), geochemical reaction rates (Hunt et al., 2015), soil depth and soil production (Hunt, 2016a; Hunt and Ghanbarian, 2016; Yu et al., 2017), and vegetation growth and soil formation (Hunt and Manzoni, 2015; Hunt, 2016b; 2017; Hunt et al., 2017).

Comparing Eq. (2) with Eq. (1) indicates that the empirical exponent $r$ can be related to the physically meaningful backbone fractal dimension (i.e., $r = 1/D_b$). Different values of $D_b$ for various percolation classes (e.g., for random and invasion percolation) are summarized in Table 2. As can be seen, within the percolation theory framework $1.217 \leq D_b \leq 1.87$ and, correspondingly, $0.535 \leq r \leq 0.826$, which agrees with the range of $r$ values reported by Alvarez (2003), among others (see Table 1). If one assumes that the traveling time follows the shortest path concepts from percolation theory, then the exponent in Eq. (2) should change to $D_{min}$, the shortest path (or minimum) fractal dimension, whose value is 1.13 or 1.374 in two or three dimensions, respectively (Porto et al., 1997). The value of 1.13 for $D_b = D_{min}$ provides a higher upper bound and thus a wider range of $r$ (i.e., $0.535 \leq r \leq 0.885$). As can be seen from Table 2, the values of $D_b$ for optimal path, site trapping invasion percolation (site TIP) and bond trapping invasion percolation (bond TIP) are not greatly different in two dimensions. All of the three percolation classes correspond to a value of about 0.82 for $r$. Similarly, bond TIP, shortest path and optimal path correspond to an $r$ value of about 0.7 in three dimensions. In addition, the values of $D_b$ for random percolation (RP) and site TIP in three dimensions in three dimensions (1.87 and 1.861, respectively), corresponding to a value of $r$ of about 0.53.

For furrow and border irrigation systems, one may normalize Eq. (2) to obtain

$$x = x_{max} \left( \frac{t}{t_{max}} \right)^{1/D_b} = \frac{x_{max}}{t_{max}^{1/D_b}} t^{1/D_b} \qquad (3)$$

where $x_{max}$ is the furrow or border length, and $t_{max}$ is the advance time to the end of the furrow or border, corresponding to the last measured data point on the advance-time curve. Comparing Eq. (3) with Eq. (1) clearly indicates that the empirical parameters $p$ and *r* can be interpreted theoretically within the percolation theory framework and have physical meaning, as will be discussed next.

**3. Materials and Methods**

In this section, we first present experimental advance-time curves and then compare them with estimates using Eqs. (1) and (3).

**3.1. Experimental data**

Experimental data used in this study are from Kamali (2015) who investigated surface flow in four open-ended furrows of the same length of 110 m, and having a longitudinal slope of 0.012 m m$^{-1}$ with rows 75 cm apart. The studied site of rectangular shape was located at the University of Tehran, Karaj, Iran. Maize (single cross 704) was planted *in-furrow* for furrows 1 and 2 and *on-ridge* for furrows 3 and 4, and irrigated once a week over a period of seven weeks. Soil texture at three depths was determined using a combination of hydrometer and sieving methods at the beginning, middle and end of the field site. Gravimetric water contents at field capacity (-33 kPa) and permanent wilting (-1500 kPa) points were measured using the pressure plate method. Salient soil properties are presented in Table 3.

Initial gravimetric water contents were measured before irrigation cycles 1 and 5. The average value for all furrows was about 8%, close to the gravimetric water content at the permanent wilting point (see Table 3). For each irrigation cycle, the advance-time was measured at eleven equally-spaced stations (i.e., at 10 m intervals) along each furrow. The inflow rate, measured with a Washington State flume (WSC) type II, was about 0.29 l/s in furrows 1 and 3, and about 0.44 l/s in furrows 2 and 4. More detailed information are presented in Table 4.

**3.2. Estimation of the advance-time curve**

To estimate the advance-time curve for surface flows using Eq. (3), one requires three parameters: $D_b$, $x_{max}$ and $t_{max}$. As noted earlier, $t_{max}$ is the advance time at the end of the furrow ($x_{max}$), corresponding to the last measured data point on the advance-time curve, which can be measured and hence is assumed to be available. Recall that $x_{max}$ = 110 m for each furrow. All four furrows had approximately the same longitudinal slope. The inflow rate in furrows 2 and 4 (i.e., 0.44 l s$^{-1}$) was 50% larger than that in furrows 1 and 3 (i.e., 0.29 l s$^{-1}$). We therefore postulate that the wetting front advance should be quasi three-dimensional in furrows 1 and 3, while quasi two-dimensional in furrows 2 and 4. Of course, water flow behind the front is inherently three-dimensional since water traverses on the surface and simultaneously infiltrates into the soil. In general, the higher the inflow rate, the faster the wetting front should move, and thus the shorter the advance time. This can be confirmed by comparing the advance time ($t_{max}$) values presented in Table 4 for all four furrows.

Surface flow over the soil-air interface involves two main processes: (1) run-on, run-off and water movement on the surface, and (2) infiltration into the soil. Since water streamlines in the run-off process have less resistance compared to water flow paths within the soil, we postulate that the optimal path class of percolation theory should accurately describe the advance-time curve. Note that the optimal path is the most energetically favorable path through a system. Since the wetting front advance in furrows 1 and 3 is probably quasi three-dimensional, and quasi two-dimensional in furrows 2 and 4, we accordingly use $D_b = 1.42$ (the optimal path fractal dimension in three dimensions; see Table 2) for furrows 1 and 3 and $D_b = 1.21$ (the optimal path fractal dimension in two dimensions; see Table 2) for furrows 2 and 4. However, some other percolation classes may well be relevant to other types of experiments depending upon such parameters as initial water content, infiltration rate, inflow rate and slope. We will discuss percolation classes and relevant $D_b$ values in the Results and Discussion section.

We further compared our results for the advance-time curve for each furrow and irrigation cycle with estimates obtained using the one-point method of Shepard et al. (1993) who recommended fixing $r$ in Eq. (1) at 0.5. Using $r = 0.5$ and the last measured data point, one has $p = x_{max}/\sqrt{t_{max}}$. Although this approach has been criticized since the value of $r$ commonly differs from 0.5, some studies showed that the Shepard et al. (1993) method still can provide reasonable appximations for furrow and border irrigation systems (e.g., Ebrahimian et al., 2010).

## 4. Results and Discussion

The measured advance-time curves of 28 experiments are shown in Figs. 3-6 for four furrows and seven irrigation cycles. Using $x_{max}$ and $t_{max}$ from the experiments, we estimated the advance-time curves using Eq. (1) with $r = 0.5$, and Eq. (3) with $D_b = 1.21$ and 1.42 for two- and three-dimensional flow patterns, respectively. As can be observed from Figs. 3-6, Eq. (3) with $D_b = 1.21$ and 1.42 (shown in red) could accurately estimate the advance-time curves for all experiments, while Eq. (1) with $r = 0.5$ and $p = x_{max}/\sqrt{t_{max}}$ (shown in blue) mostly underestimates the curves.

We found that the optimal path class of percolation theory with $D_b$ = 1.21 (for two dimensions) and 1.42 (for three dimensions), corresponding respectively to $r$ = 0.826 and 0.704 in Eq. (1) (see Table 2), accurately estimated the advance-time curves for 28 experiments carried out by Kamali (2015). Still, one should note that other percolation classes might provide more accurate estimates for other experiments. The value of $D_b$ in Eq. (3), in fact, depends on several factors, such as slope, inflow rate, initial water content, and infiltration rate. The steeper the slope, the greater the influence of gravity and run-off, and thus the shorter the advance time. This means that in furrows with mild slopes a higher portion of water at the wetting front would infiltrate into the soil while advancing through the furrow as compared to furrow having steep slopes (e.g., Esfandiari and Maheshvari, 1997). This implies that water flow at the wetting front should be quasi three- and two-dimensional in mild and steep furrows, respectively.

The inflow rate has an influence on the advance-time curve similar to the slope. The higher the inflow rate, the less water will infiltrate at the front, thus making the wetting front advance quasi two-dimensional along the furrow, as we showed for furrows 2 and 4. This is the main reason why $D_b$ = 1.21 (for two dimensions) could accurately describe the advance-time curves for furrows 2 and 4 with inflow rates of about 0.44 l s$^{-1}$ while $D_b$ = 1.42 (for three dimensions) did hold in furrows 1 and 3 having inflow rates of near 0.29 l s$^{-1}$.

The initial water content of soil prior to the irrigation event also affects the wetting front advance since it controls infiltration and hence also run-off. The higher the initial water content, the lower the infiltration rate, and thus the faster the water will move along the furrow (Rasoulzadeh and Sepaskhah, 2003). Less infiltration into the soil at the wetting front in turn will cause water to traverse more quasi two-dimensional.

The rate of infiltration from the furrow into the soil is very much controlled by the hydraulic conductivity of the soil, and hence texture. Fine-textured soils with their lower hydraulic conductivities hence will show lower infiltration rates. For this reason, one should expect the wetting front advance to be more three-dimensional in coarse-textured and more two-dimensional in fine-textured soils. However, if cracks exist along the furrow surface, particularly in fine-textured soils, preferential flow paths will be created (e.g., Novak et al., 2000) that will likely cause the advancing wetting front in the furrow to become quasi three-dimensional. This still has to be demonstrated through either experiments or numerical simulations, including how such on-site field conditions may change the values of $r$ and $D_b$ in the advance-time curves. Interestingly, planting maize in furrows or on ridges did not substantially affect the exponent $r$ in Eq. (1) or $D_b$ in Eq. (3). As we showed in Figs. 3-6, the same $D_b$ values used to estimate the advance-time curves for furrows in which maize was planted, did hold for furrows in which maize was planted on ridges.

A comparison of the range ($0.48 \leq r \leq 0.68$) and average value (0.58) of $r$ for border irrigation with those for furrow irrigation in Table 1 indicates that one should expect a smaller value of $r$ for borders than that for furrows. We, accordingly, postulate that the wetting front advance should follow a quasi three-dimensional pattern in borders, while being more quasi two-dimensional in furrows. This means that the two-dimensional values of $D_b$, shown in Table 2, should be relevant for furrows, and the three-dimensional values for borders.

We also postulate that for closely-spaced furrows with steep slopes and relatively high inflow rates, the wetting front advance will be mostly quasi one-dimensional. Since the backbone fractal dimension in one dimension is 1, one should hence expect the exponent $r$ in Eq. (1) to be close to 1 ($r = 1/D_b$). This would then be in agreement with experimental observations by Khatri and Smith (2006) who reported $r$ values as large as 0.97 (see Table 1).

Although we showed that the optimal path percolation class could estimate precisely the advance-time curves, one should not expect percolation theory, as developed mostly to study subsurface flow, to accurately describe surface flow for all types of experiments. In fact, additional and more comprehensive sets of experiments are required to investigate the effects of initial water content, inflow rate, Manning's roughness coefficient, infiltration rate, soil texture and surface slope on the exponent $r$, and to find out which percolation class and $D_b$ value would provide more accurate estimates under various field conditions.

## 5. Conclusions

In this study, we proposed theoretical bounds for the exponent $r$ in the empirical power-law advance-time curve for surface flow. We showed that the exponent $r$ can be linked to the backbone fractal dimension ($D_b$) of the percolating cluster within the percolation theory framework ($r = 1/D_b$). A comparison with $r$ values reported in the literature showed that the theoretical range of the inverse of $D_b$ (i.e., $0.535 \leq 1/D_b \leq 0.885$) is in well agreement with the experimental range found for furrow and border irrigation systems. We also showed that the optimal path class from percolation theory could accurately estimate the advance-time curves for 28 furrow experiments, while the one-point method of Shepard et al. (1993) mostly underestimated the curves. We found that large and small values of $r$ should be respectively relevant to quasi two- and three-dimensional wetting front advance.

**Acknowledgement**

BG acknowledges supports through faculty startup funds from Kansas State University.

Ebrahimian, H. (2014). Soil infiltration characteristics in alternate and conventional furrow irrigation using different estimation methods. Korean Society of Civil Engineers. 18(6), 1904-1911.

Elliott RL, Walker WR (1982) Field evaluation of furrow infiltration and advance functions. Trans ASAE 25(2), 396-400.

Esfandiari, M., & Maheshwari, B. L. (1997). Field values of the shape factor for estimating surface storage in furrows on a clay soil. Irrigation Science, 17(4), 157-161.

Feder, J. (1988), Fractals, Plenum, New York.

Gefen, Y., Aharony, A., & Alexander, S. (1983). Anomalous diffusion on percolating clusters. Physical Review Letters, 50(1), 77-80.

Hunt, A. G. (2015). Predicting rates of weathering rind formation. Vadose Zone Journal, 14(7).

Hunt, A. G. (2016a). Soil depth and soil production. Complexity, 21(6), 42-49.

Hunt, A. G. (2016b). Spatio-temporal scaling of vegetation growth and soil formation from percolation theory. Vadose Zone Journal, 15(2).

Hunt, A. G. (2017). Spatiotemporal Scaling of Vegetation Growth and Soil Formation: Explicit Predictions. Vadose Zone Journal, 16(6).

Hunt, A. G., and S. Manzoni (2015), Networks on Networks: The Physics of Geobiology and Geochemistry, Inst. of Phys., Bristol, U.K.

Hunt, A. G., & Ghanbarian, B. (2016). Percolation theory for solute transport in porous media: Geochemistry, geomorphology, and carbon cycling. Water Resources Research, 52(9), 7444-7459.
16

Table 1. Values of the exponent *r* in Eq. (1) reported in the literature for various conditions.

| Reference | Irrigation system | No. of samples | *r* | Ave. *r* | Remarks |
|---|---|---|---|---|---|
| Alvarez (2003) | Furrow | 12 | 0.58-0.72 | 0.66 | From various references, see their Table 1 |
| Abbasi et al. (2003) | Furrow | 3 | 0.62-0.91 | 0.76 | Maricopa Agricultural Center, Phoenix AZ |
| Khatri and Smith (2006) | Furrow | 27 | 0.73-0.97 | 0.86 | Cotton field T, Southern Queensland |
|  | Furrow | 17 | 0.62-0.85 | 0.72 | Cotton field C, Southern Queensland |
| Bautista et al. (2009) | Furrow | 4 | 0.57-0.81 | 0.67 | Four furrows of various lengths and soil textures |
| Abbasi et al. (2009a) | Furrow | 3 | 0.69-0.91 | 0.77 | Seed & Plant Improvement Res. Inst., Karaj, Iran |
| Abbasi et al. (2009b) | Furrow | 6 | 0.62-0.91 | 0.76 | Research Station for Tobacco, Urmia, Iran |
| Ebrahimian et al. (2010) | Furrow | 5 | 0.52-0.77 | 0.63 | From various references, see their Table 1 |
|  | Border | 6 | 0.48-0.68 | 0.58 | From various references, see their Table 2 |
| Ebrahimian (2014) | CFI[*] | 7 | 0.65-0.86 | 0.76 | Maize field, Karaj, Iran |
|  | FFI | 7 | 0.65-0.70 | 0.68 | Maize field, Karaj, Iran |
|  | AFI | 7 | 0.63-0.75 | 0.69 | Maize field, Karaj, Iran |

[*] CFI: conventional furrow irrigation, FFI: fixed alternate furrow irrigation, AFI: variable alternate furrow irrigation.

Table 2. Theoretical $D_b$ and corresponding $r$ values for various percolation classes in two and three dimensions (Sheppard et al., 1999).

| Dimension | Percolation class | $D_b$ | $r$ (=1/$D_b$) |
|---|---|---|---|
| 2 | RP* | 1.643 | 0.609 |
| 2 | Optimal path | 1.210 | 0.826 |
| 2 | Site TIP | 1.217 | 0.822 |
| 2 | Bond TIP | 1.217 | 0.822 |
| 2 | Shortest path | 1.130† | 0.885 |
| 3 | RP | 1.870 | 0.535 |
| 3 | Optimal path | 1.420 | 0.704 |
| 3 | Site TIP | 1.861 | 0.537 |
| 3 | Bond TIP | 1.458 | 0.686 |
| 3 | Shortest path | 1.376† | 0.727 |

* RP is random percolation and TIP is trapping invasion percolation
† 1.130 and 1.376 represent shortest path (or minimum) fractal dimensions

Table 3. Salient soil properties of the field site studied by Kamali (2015).

| Location | Depth (m) | Texture | Clay (%) | Silt (%) | Sand (%) | BD* (g cm$^{-3}$) | FC (%) | PWP (%) | EC (dS m$^{-1}$) |
|---|---|---|---|---|---|---|---|---|---|
| Beginning | 0-0.2 | Clay loam | 28.5 | 35 | 36.5 | 1.5 | 18.2 | 8.7 | 2.55 |
|  | 0.2-0.4 | Clay loam | 28.5 | 33.8 | 37.8 | 1.45 | 17.5 | 8.1 | 1.77 |
|  | 0.4-0.6 | Sandy loam | 16 | 17.5 | 66.5 | 1.47 | 14.2 | 6 | 1.88 |
| Middle | 0-0.2 | Loam | 26 | 30 | 44 | 1.5 | 18.1 | 8.5 | 2.1 |
|  | 0.2-0.4 | Sandy clay loam | 23.5 | 25 | 51.5 | 1.45 | 17.2 | 8 | 2.08 |
|  | 0.4-0.6 | Sandy clay loam | 21 | 22.5 | 56.5 | 1.52 | 15.5 | 6.9 | 2 |
| End | 0-0.2 | Clay loam | 31 | 31.7 | 37.3 | 1.51 | 18.1 | 8.4 | 2.98 |
|  | 0.2-0.4 | Loam | 26.8 | 30.4 | 42.8 | 1.48 | 17.7 | 8.1 | 2.05 |
|  | 0.4-0.6 | Sandy loam | 20.2 | 24.6 | 55.3 | 1.49 | 15 | 6.6 | 2.47 |

* BD is bulk density, FC is gravimetric water content at field capacity (-33 kPa matric potential), PWP is gravimetric water content at permanent wilting point (-1500 kPa matric potential), EC is electrical conductivity of saturated soil sample

Table 4. Properties of the investigated irrigation scheme with seven cycles irrigating four furrows.

| Furrow | Maize planted | Parameter | Irrigation cycle | | | | | | |
|---|---|---|---|---|---|---|---|---|---|
| | | | 1 | 2 | 3 | 4 | 5 | 6 | 7 |
| 1 | in furrow | inflow rate (l s$^{-1}$) | 0.32 | 0.31 | 0.29 | 0.28 | 0.28 | 0.29 | 0.29 |
| | | run-off volume (l) | 1696 | 1837 | 1202 | 1067 | 1093 | 1433 | 878 |
| | | infiltration volume (l) | 2670 | 2450 | 2688 | 2179 | 2548 | 1980 | 1921 |
| | | cut-off time (min) | 229 | 245 | 230 | 200 | 225 | 220 | 165 |
| | | $t_{max}$* (min) | 51 | 59.5 | 73 | 61.5 | 67.5 | 71 | 63 |
| 2 | in furrow | inflow rate (l s$^{-1}$) | 0.44 | 0.44 | 0.44 | 0.46 | 0.44 | 0.44 | 0.44 |
| | | run-off volume (l) | 2634 | 3104 | 2721 | 1907 | 2255 | 2652 | 1719 |
| | | infiltration volume (l) | 3252 | 2577 | 2521 | 2084 | 2413 | 2561 | 2033 |
| | | cut-off time (min) | 229 | 220 | 205 | 150 | 180 | 200 | 145 |
| | | $t_{max}$ (min) | 45.5 | 44 | 44.6 | 39.5 | 39.5 | 40 | 41 |
| 3 | on ridge | inflow rate (l s$^{-1}$) | 0.34 | 0.31 | 0.29 | 0.28 | 0.29 | 0.29 | 0.29 |
| | | run-off volume (l) | 1720 | 1934 | 1316 | 1369 | 1318 | 1232 | 953 |
| | | infiltration volume (l) | 2440 | 2833 | 2584 | 2296 | 2079 | 1907 | 1837 |
| | | cut-off time (min) | 204 | 265 | 230 | 225 | 200 | 180 | 165 |
| | | $t_{max}$ (min) | 45 | 95 | 76 | 77.5 | 60.6 | 58.7 | 61.5 |
| 4 | on ridge | inflow rate (l s$^{-1}$) | 0.44 | 0.44 | 0.44 | 0.46 | 0.44 | 0.44 | 0.44 |
| | | run-off volume (l) | 2405 | 3043 | 2184 | 2136 | 2381 | 2113 | 1601 |
| | | infiltration volume (l) | 2875 | 2641 | 3115 | 2570 | 2800 | 2567 | 2122 |
| | | cut-off time (min) | 204 | 265 | 230 | 225 | 200 | 180 | 165 |
| | | $t_{max}$ (min) | 37 | 42.6 | 53.5 | 50 | 49.9 | 41 | 43.5 |

* Advance time at the end of furrow.

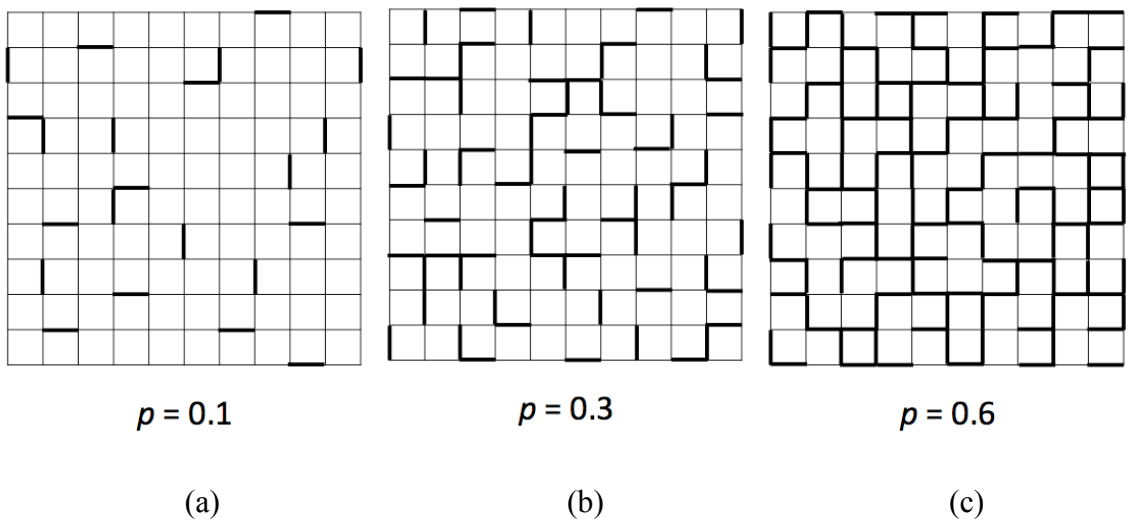

Figure 1. Bond percolation on a square lattice with occupation probability $p = 0.1$ (a), 0.3 (b) and 0.6 (c). The critical probability $p_c$ on the square lattice is 0.5 for bond percolation in two dimensions.

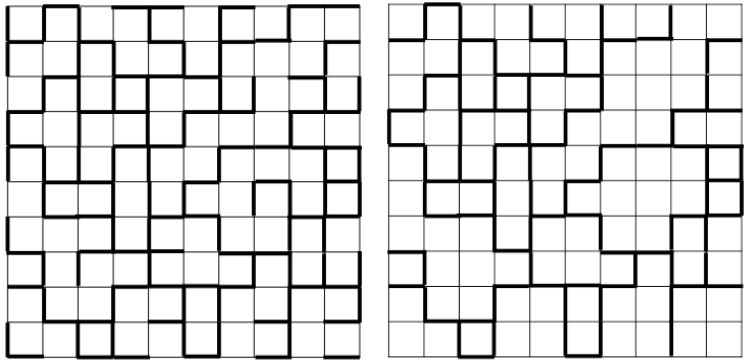

(a) (b)

Figure 2. (a) Bond percolation on a square lattice with occupation probability $p = 0.6$ and (b) its backbone. The value of the backbone fractal dimension $D_b$ for random and bond trapping invasion percolation is, respectively, 1.643 and 1.217 (Sheppard et al., 1999). The values of $D_b$ for different percolation classes are listed in Table 2.

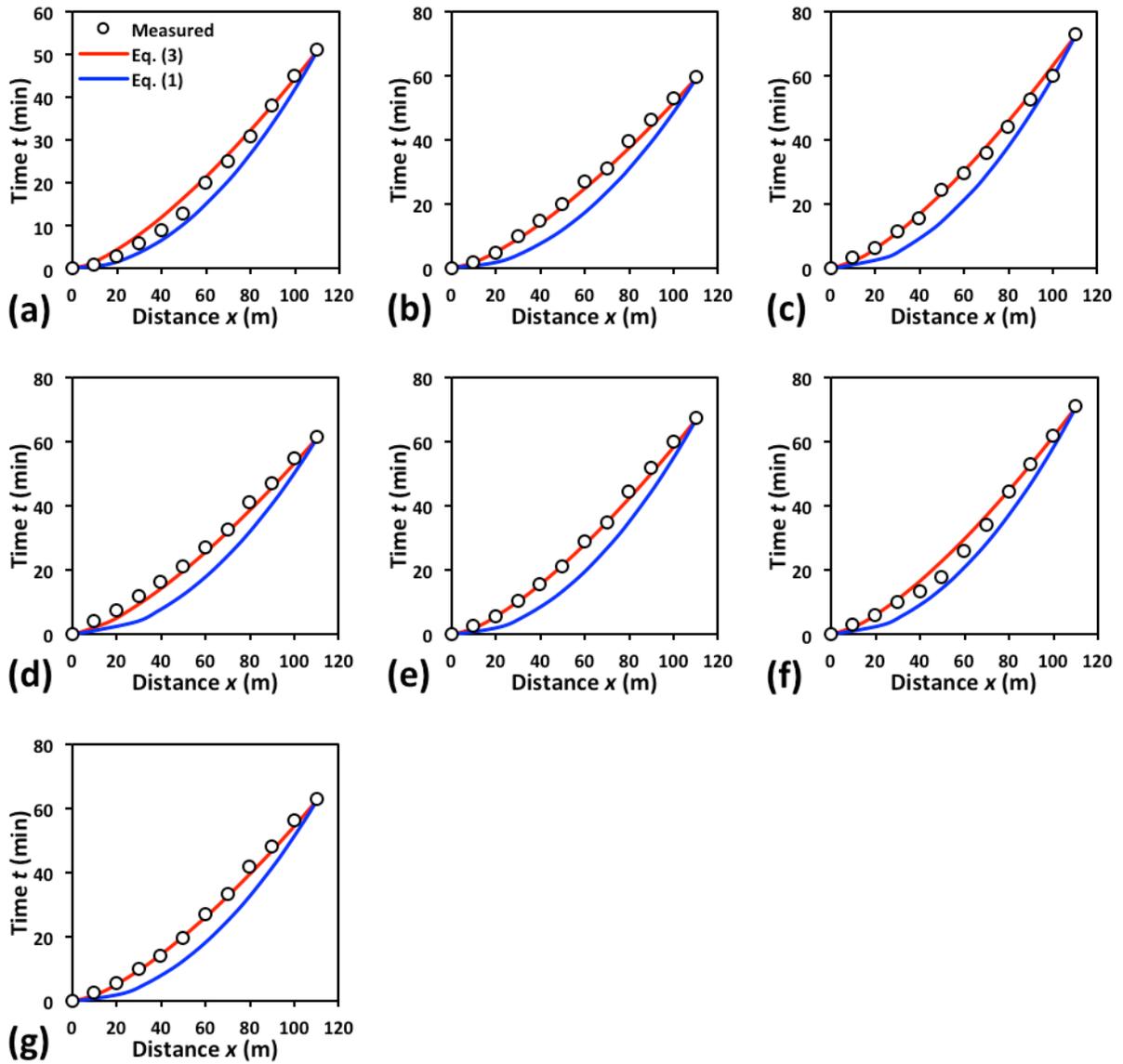

Figure 3. Measured advance-time curves, as well as the estimated curves using Eq. (1) with $r = 0.5$ and $p = x_{max}/\sqrt{t_{max}}$ (Shepard et al., 1993), represented by blue curves, and Eq. (3) with $D_b = 1.42$ (optimal path in three dimensions; see Table 2), denoted by red curves, using the advance time ($t_{max}$) values measured at the end of furrow 1 for irrigation cycles 1 to 7 (a-g).

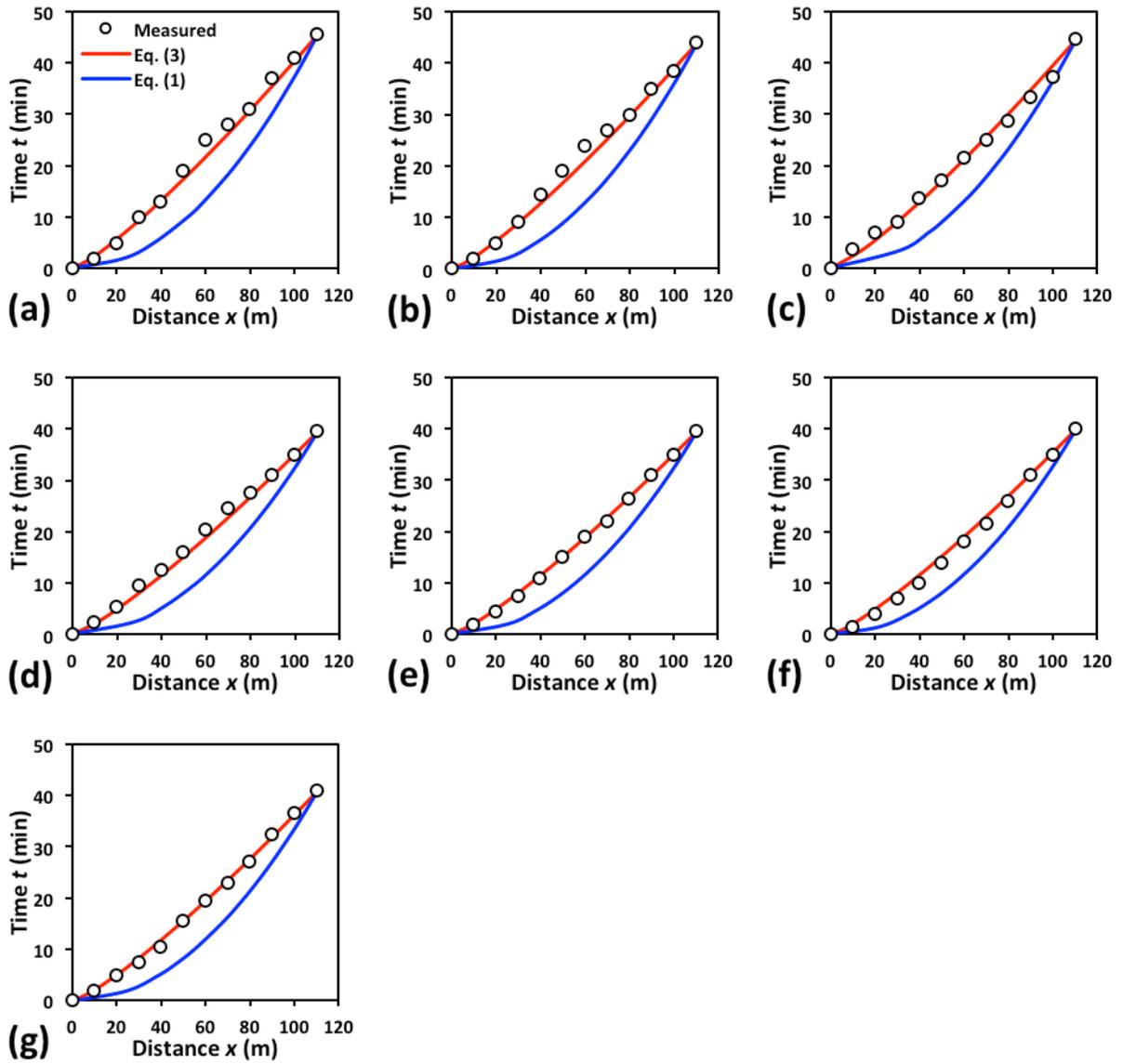

Figure 4. Measured advance-time curves, as well as the estimated curves using Eq. (1) with $r =$ 0.5 and $p = x_{max}/\sqrt{t_{max}}$ (Shepard et al., 1993), represented by blue curves, and Eq. (3) with $D_b =$ 1.21 (optimal path in two dimensions; see Table 2), denoted by red curves, using the advance time ($t_{max}$) values measured at the end of furrow 2 for irrigation cycles 1 to 7 (a-g).

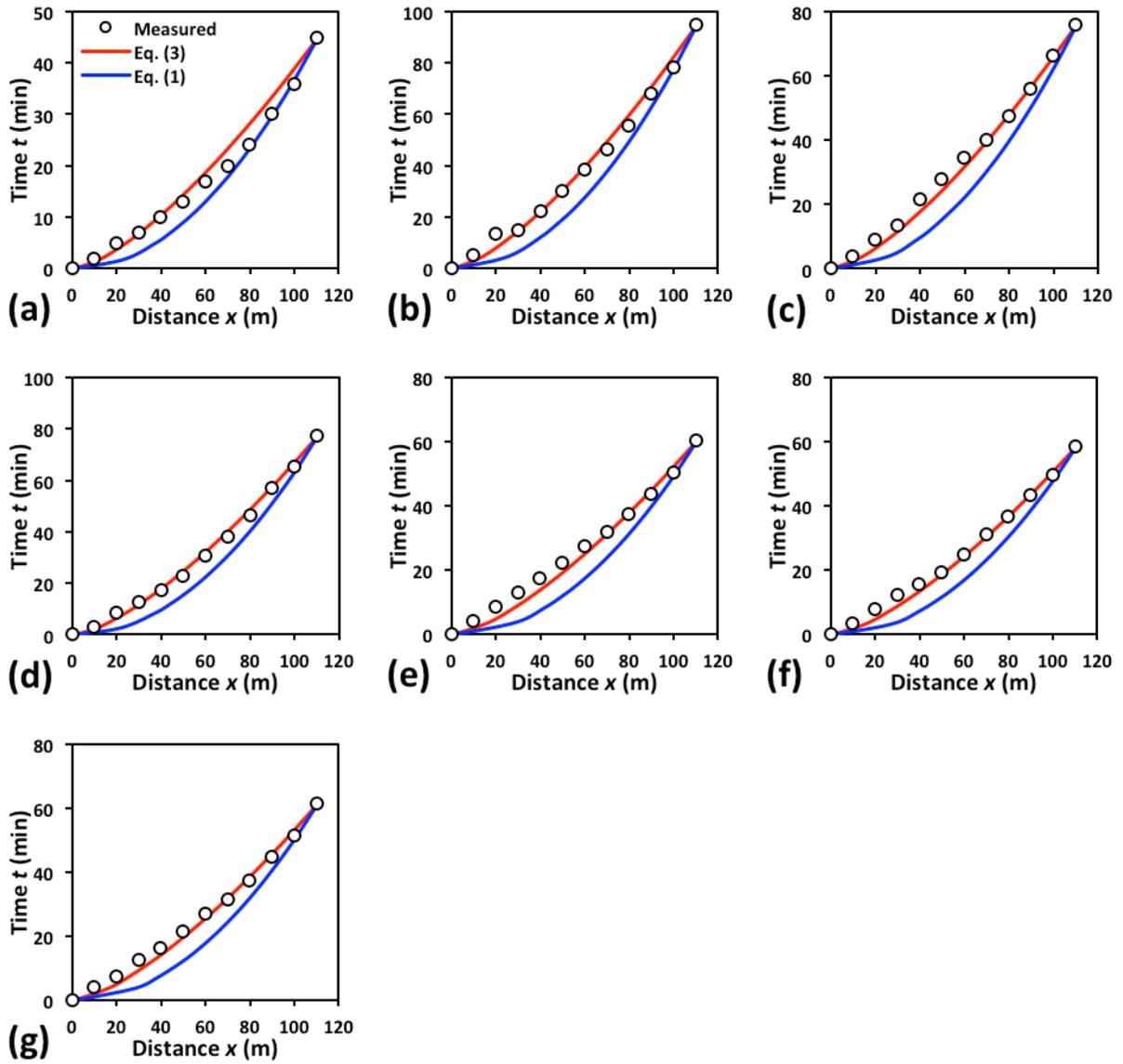

Figure 5. Measured advance-time curves, as well as the estimated curves using Eq. (1) with $r =$ 0.5 and $p = x_{max}/\sqrt{t_{max}}$ (Shepard et al., 1993), represented by blue curves, and Eq. (3) with $D_b =$ 1.42 (optimal path in three dimensions; see Table 2), denoted by red curves, using the advance time ($t_{max}$) values measured at the end of furrow 3 for irrigation cycles 1 to 7 (a-g).

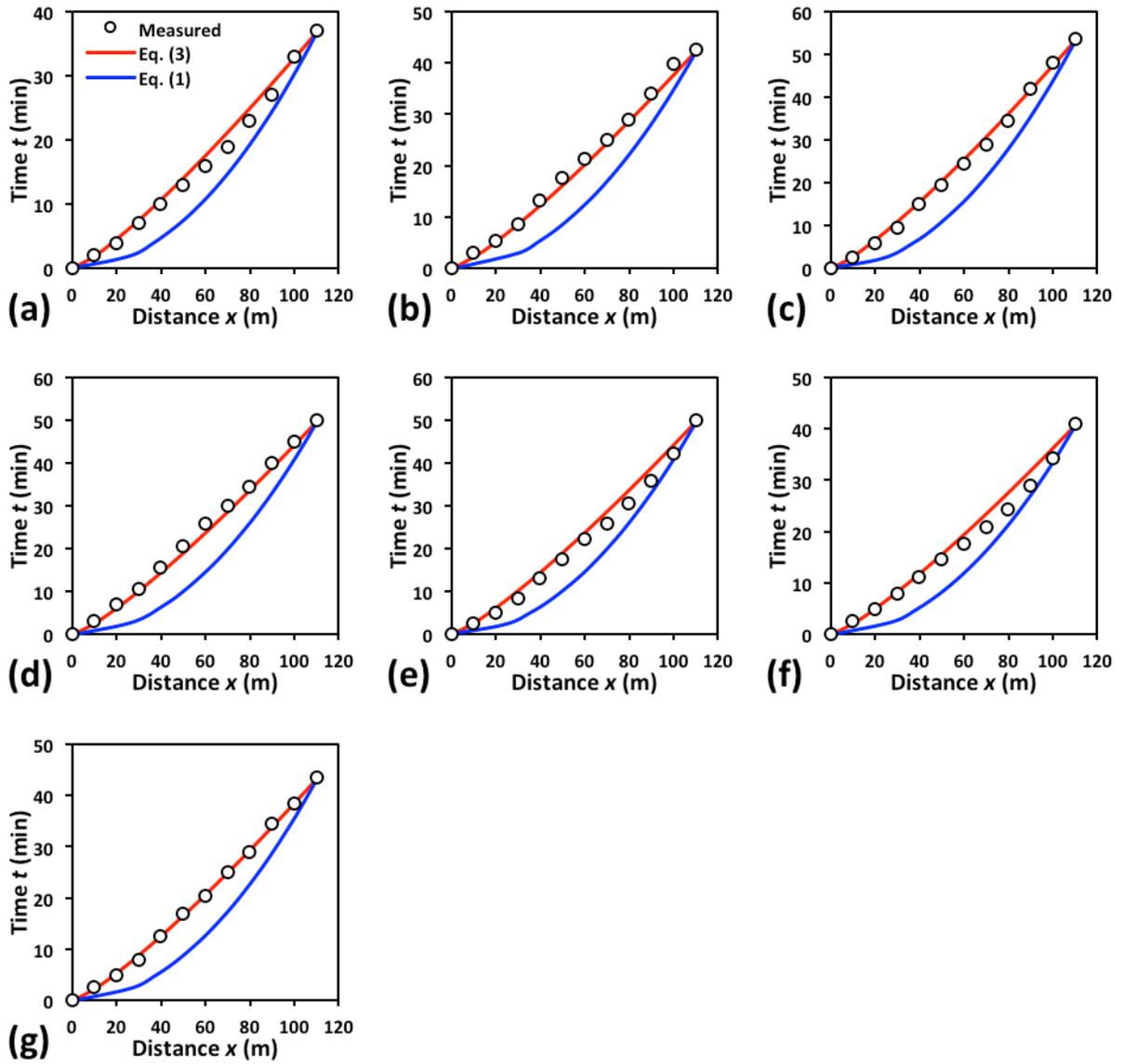

Figure 4. Measured advance-time curves, as well as the estimated curves using Eq. (1) with $r = 0.5$ and $p = x_{max}/\sqrt{t_{max}}$ (Shepard et al., 1993), represented by blue curves, and Eq. (3) with $D_b = 1.21$ (optimal path in two dimensions; see Table 2), denoted by red curves, using the advance time ($t_{max}$) values measured at the end of furrow 4 for irrigation cycles 1 to 7 (a-g).